\documentclass[english,aps,manuscript]{revtex4}
\usepackage{mathptmx}
\usepackage[T1]{fontenc}
\usepackage[latin9]{inputenc}
\usepackage{graphicx}

\makeatletter

\DeclareRobustCommand{\greektext}{%
  \fontencoding{LGR}\selectfont\def\encodingdefault{LGR}}
\DeclareRobustCommand{\textgreek}[1]{\leavevmode{%
  \IfFileExists{grtm10.tfm}{}{\fontfamily{cmr}}\greektext #1}}
\DeclareFontEncoding{LGR}{}{}
\DeclareTextSymbol{\~}{LGR}{126}

\@ifundefined{textcolor}{}
{%
 \definecolor{BLACK}{gray}{0}
 \definecolor{WHITE}{gray}{1}
 \definecolor{RED}{rgb}{1,0,0}
 \definecolor{GREEN}{rgb}{0,1,0}
 \definecolor{BLUE}{rgb}{0,0,1}
 \definecolor{CYAN}{cmyk}{1,0,0,0}
 \definecolor{MAGENTA}{cmyk}{0,1,0,0}
 \definecolor{YELLOW}{cmyk}{0,0,1,0}
 }

\makeatother

\usepackage{babel}

\begin{document}

\title{Improving the Quality Factor of Microwave Compact Resonators by Optimizing
their Geometrical Parameters}

\author{K. Geerlings, S. Shankar, E. Edwards, L. Frunzio, R.J. Schoelkopf,
M.H. Devoret}

\address{Department of Applied Physics, Yale University, New Haven, Connecticut
06520-8284, USA}
\begin{abstract}
Applications in quantum information processing and photon detectors are stimulating a race to produce the highest possible quality factor on-chip superconducting microwave resonators. We have tested the surface-dominated loss hypothesis by systematically studying the role of geometrical parameters on the internal quality factors of compact resonators patterned in Nb on sapphire. Their single-photon internal quality factors were found to increase with the distance between capacitor fingers, the width of the capacitor fingers, and the resonator impedance. Quality factors were improved from 210,000 to 500,000 at T = 200 mK. All of these results are consistent with our starting hypothesis.
\end{abstract}
\maketitle
Improving the internal quality factor of on-chip microwave superconducting
resonators is a key development for quantum information processing
and photon detectors \cite{Wang2011,Mazin2006}. The internal quality
factor at single-photon powers, $Q{}_{i}$, of particular interest
for quantum information applications, is observed to be 10-100 times
lower than high-power quality factors \cite{Lindstroem2009,Vissers2010a,aBarends2010,Khalil2011,Weber2011a}.
Ideas for increasing resonator $Q{}_{i}$ include switching from conventional
metals like Nb or Al to alloys such as TiN or NbTiN \cite{Barends2008,Vissers2010a,Leduc2010a,aBarends2010},
using interface layers of SiN \cite{Vissers2010a}, etching the substrate
between traces \cite{aBarends2010}, depositing metal under special
conditions \cite{Megrant2012}, or using low loss substrates \cite{Weber2011a}.
Results from these experiments have generated the hypothesis that
resonator $Q{}_{i}$ is limited by a surface two level system (TLS)
distribution \cite{Gao2008,Barends2008,Khalil2011,aBarends2010}.

\begin{figure}
\begin{centering}
\includegraphics{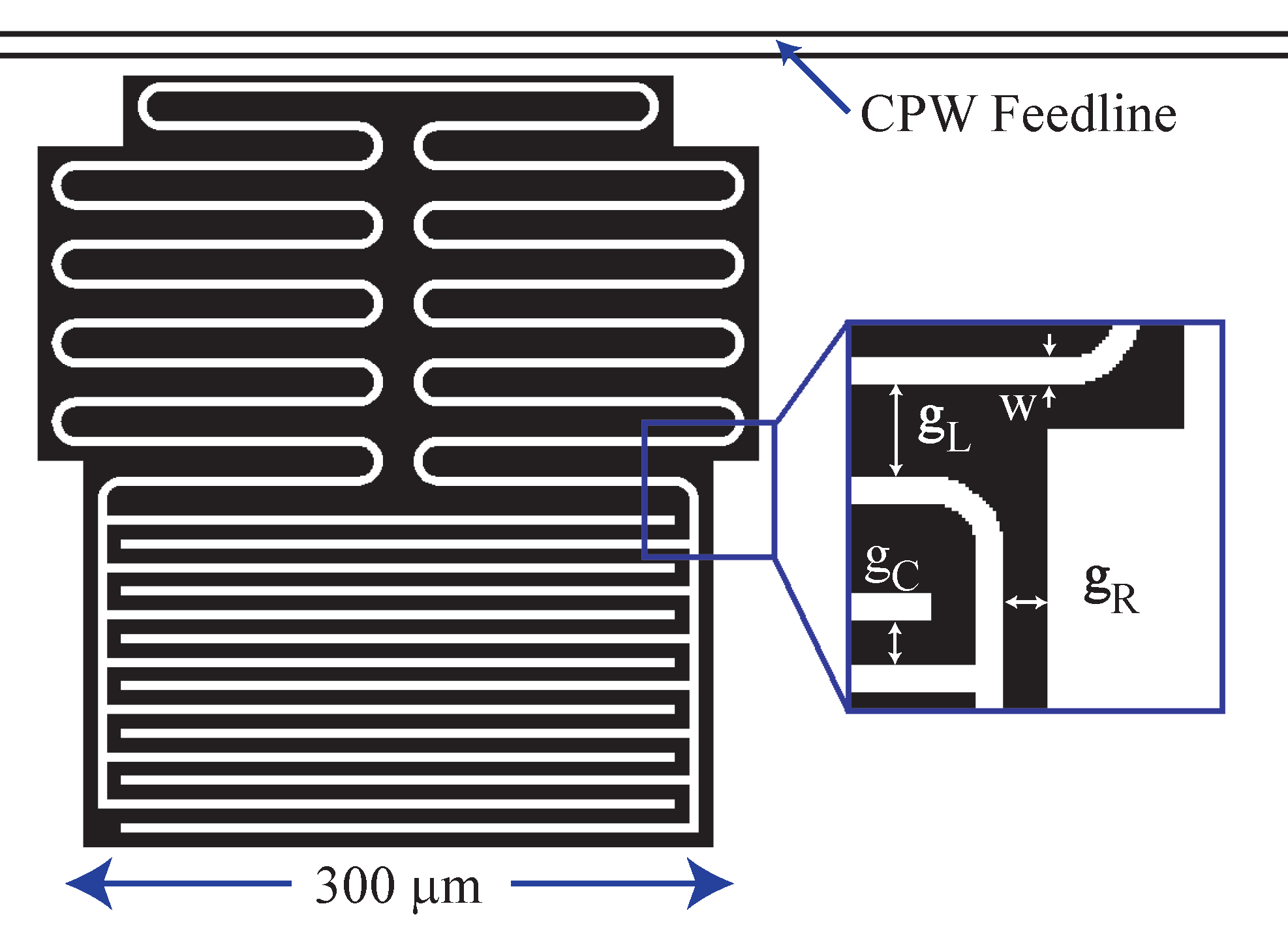}
\par\end{centering}

\caption{\label{fig:compact resonator schematic}Schematic of compact resonator
and inset showing resonator parameters. The compact resonator is coupled
inductively to the CPW feedline. The parameters indicated in the inset directly affect the participation of the insulator and metal surfaces to the reactive elements of the resonator.}
\end{figure}

Motivated by a previous study that showed that the $Q{}_{i}$ of coplanar
waveguide (CPW) resonators increases with increasing gap \cite{Gao2008},
we extended the idea of geometrical optimization to compact resonators \cite{Khalil2011,Lindstroem2009,Leduc2010a}.
Compact resonators, as shown in Figure \ref{fig:compact resonator schematic},
consist of a meander inductor in parallel to an interdigitated capacitor.
Their small size makes them an ideal element for multi-qubit processors.
While compact resonators have been shown to have similar $Q{}_{i}$
as the more widely used CPW resonators \cite{Khalil2011}, they permit
more design choices. Here we show that by changing parameters linked to the surface participation ratio, we have
optimized these resonators to achieve an improvement by a factor of 2.4 $\pm$ 0.2.  We have thus been able to reach a $Q{}_{i}$ of 500,000 at a resonator temperature of 200 mK, our point of reference.  In this paper, we prefer to quote $Q{}_{i}$ at this temperature because we believe that even when the sample box is anchored to a colder plate, resonator temperatures substantially below 200 mK may not be reached reliably. We return to this point later in the paper.

\begin{figure}
\begin{centering}
\includegraphics{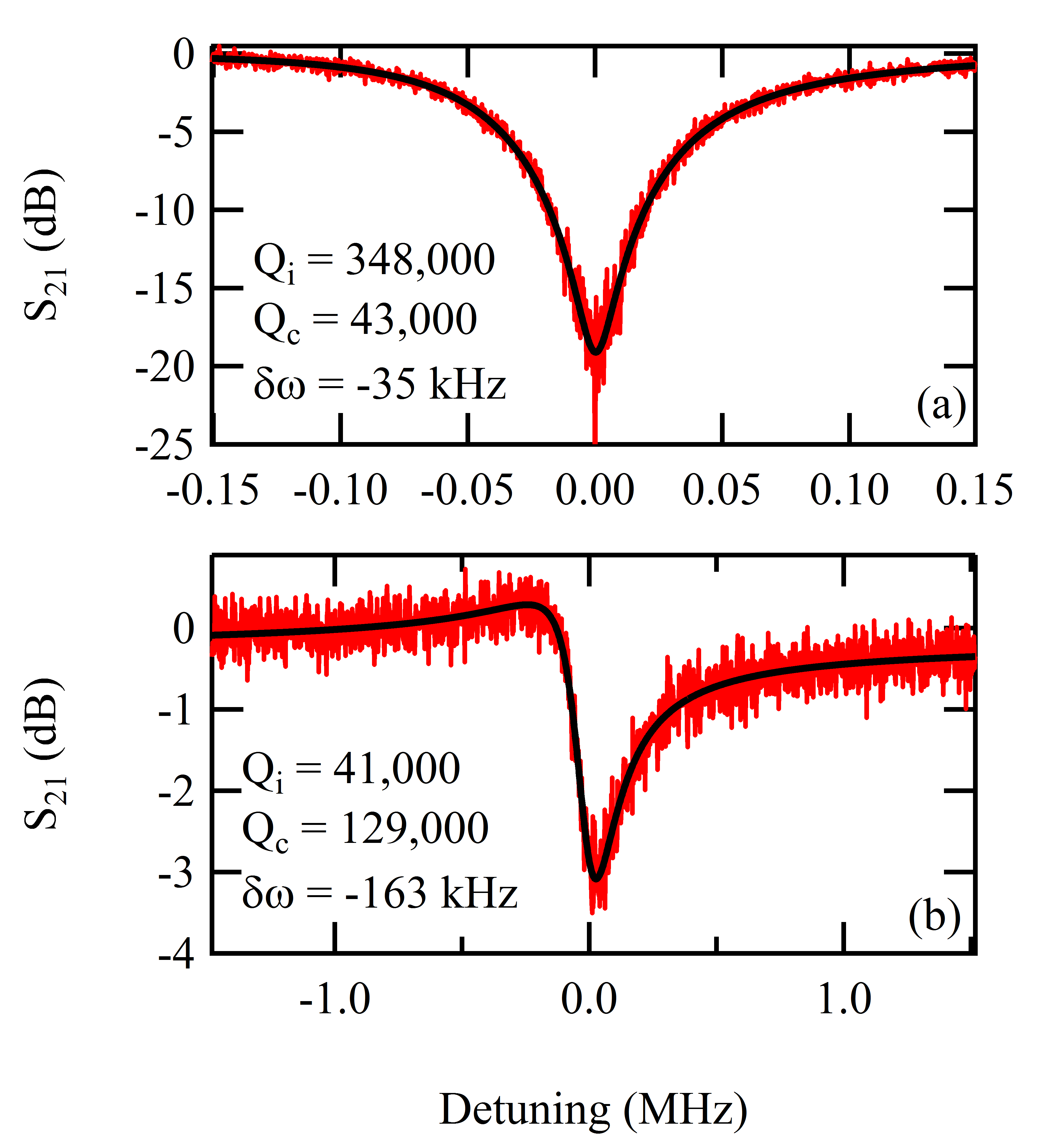}
\par\end{centering}

\caption{Two extreme examples of resonator response curves fit with Equation
\ref{eq:fitting expression}. Responses typically fall in between
(a) symmetric, and (b) strongly asymmetric about the resonant frequency.\label{fig:example fits}}
\end{figure}

We measure the quality factor of our compact resonators by performing
a microwave transmission experiment. Coupling to the resonators is
achieved by placing the resonator in a cutout in the ground plane
of a CPW feedline, relying on the mutual inductance between the feedline
and the resonator inductor. This coupling introduces a second quality
factor, the coupling quality factor $Q{}_{c}$. Typical values of
$Q{}_{c}$ that we designed ranged from 20,000 to 150,000. As a control
experiment, we have designed and measured resonators with $Q{}_{c}$
as high as $1.6\times10{}^{6}$ with no change in $Q{}_{i}$. As shown
in \cite{Megrant2012,Khalil2012}, the measurement of microwave transmission
$S{}_{21}$ through the feedline as a function of frequency $\omega$
provides access to $Q{}_{i}$. Although simple resonator models predict
a symmetric $S{}_{21}$ response, the measured response is typically
asymmetric due to reflections in the feedline circuit, as shown in
Figure \ref{fig:example fits}. Nevertheless, the theory of the arbitrary linear circuit model with one pole and perfect transmission at zero frequency shows that the asymmetric response can still be
fit to separately extract $Q{}_{c}$ and $Q{}_{i}$ by introducing
an extra parameter $\delta\omega$ characterizing the asymmetry. We thus analyze our data with
Equation \ref{eq:fitting expression}, where the total quality factor,
$Q{}_{0}$, is defined as $1/Q_{0}=1/Q_{i}+1/Q_{c}$.

\begin{equation}
S_{21}=1-\frac{\frac{Q_{0}}{Q_{c}}-2iQ_{0}\frac{\delta\omega}{\omega_{0}}}{1+2iQ_{0}\frac{\omega-\omega_{0}}{\omega_{0}}}\label{eq:fitting expression}\end{equation}

This expression is exactly equivalent to Eq. (13) in \cite{Khalil2012} and to Eq. (3) in \cite{Megrant2012} with a different parametrization.

In our measurement setup, we cool 4 chips at once in a dilution refrigerator
with a base temperature of 80 mK (sample temperature of approximately
200 mK). Since each of our chips contained
one feedline coupled to 6 independent resonators at frequencies between
5 and 8 GHz, 24 resonators were tested in each cooldown. The chips
were wire-bonded to a printed circuit board with Arlon dielectric
and placed inside a copper sample box. The box was mounted inside
a magnetic shield (Amuneal A4K) and attenuators were installed totaling
50 dB on the input microwave line. All four chips were excited simultaneously
using a passive 4-way microwave splitter. The output line consisted
of two Pamtech 4-8 GHz isolators on the mixing chamber, a 12 GHz low-pass
filter on the 700 mK stage, and a Caltech HEMT amplifier at the 4
K stage. The measurement line was switched between the 4 chips using
a microwave switch (Radiall R573423600) mounted on the mixing chamber.

Our resonators were fabricated using etched Nb on c-plane sapphire.
Before metal deposition, the sapphire surface was prepared by a 60
s ion-milling using a 3 cm Kaufmann source that shoots 500 eV Argon
ions at our wafer. Our source operates at a flow rate of 4.25 sccm
and a pressure of about 10 $\mu$Torr, generating a current
density of 0.67 mA/cm$^{2}$. A 200 nm layer of Nb was then dc magnetron sputtered
on the wafer. Photolithography was performed by patterning directly
onto S1808 resist using a 365 nm laser. After development, the Nb
was etched using a 1:2 mixture of Ar:$\mathrm{SF}{}_{6}$ at 10 mTorr
for 3 minutes. The wafer was then diced into individual chips for
measurement.

In the systematic variation of compact resonator parameters, we chose
to optimize the following parameters shown pictorially in Figure \ref{fig:compact resonator schematic};
the gap $\mathit{g_{C}}$ between two adjacent capacitor fingers,
the distance $g{}_{L}$ between two adjacent inductor meanders, the
distance $g{}_{R}$ between the resonator and the surrounding ground
plane, and the width $w$ of the resonator traces. In addition, we
also varied the characteristic impedance $Z{}_{0}$ of the resonator.
This set of parameters is relevant for surface losses.

We formed a benchmark set of resonators with parameter values: $g{}_{C}$
= 10 $\mu$m, $g{}_{L}$ = 20 $\mu$m, $g_{R}$ = 10
$\mu$m, $w$ = 5 $\mu$m and $Z{}_{0}$ = 100 \textgreek{W}.
Resonators with this set of parameters will now be called {}``Design
A'' resonators. We measured 25 Design A resonators with an average
$Q{}_{i}$ of 160,000($\pm$20,000) and a maximum of 210,000 at single-photon
power. Additionally, one chip with 6 resonators inexplicably had quality
factors ranging from 40,000 to 70,000, much lower than the rest; we
did not include this chip in the benchmark. $Q{}_{i}$ typically increased
to around $1\times10{}^{6}$ at a {}``high'' power corresponding
to an average of $10{}^{8}$ photons in the resonator. The resonant
frequency typically decreased as the temperature passed below 1.3$\:$K,
consistent with TLS loss \cite{Gao2008}. These results are consistent
with the hypothesis that our benchmark $Q{}_{i}$ is controlled by
surface losses.

\begin{figure}
\begin{centering}
\includegraphics[width=3.375in]{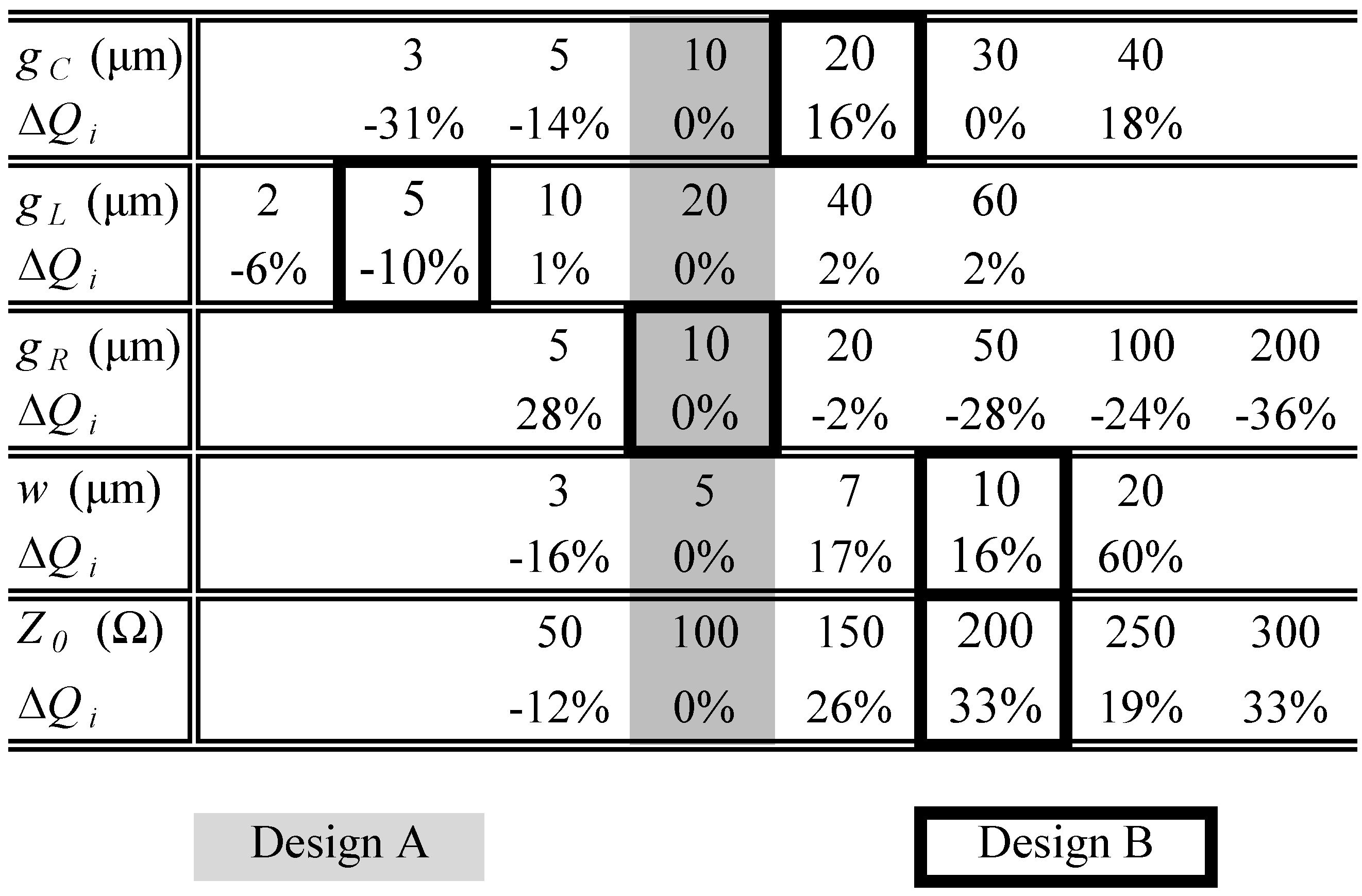}
\par\end{centering}

\caption{\label{fig:parameter results table}Dependence of $Q{}_{i}$ on parameter
values. The changes in $Q{}_{i}$ for a given mutant value are reported
in reference to the average $Q{}_{i}$ (160,000) of Design A with
positive values representing improvements. The shaded column indicates
the Design A value of that parameter; $\Delta Q{}_{i}$ in this column
is zero by definition. The square in each row with a bold border shows
the value chosen for Design B. Parameters for Design C cannot easily
be shown on this figure as explained in the main text.}
\end{figure}

We measured 24 geometrical variants of Design A, with each {}``mutant''
resonator having only one parameter value that is changed. For example,
the mutant values of $g{}_{C}$ were: 3, 5, 20, 30, and 40 $\mu$m.
The results of the mutant resonators are shown in Figure \ref{fig:parameter results table};
percent changes in $Q{}_{i}$ are given with respect to the Design
A resonator benchmark.

For $g{}_{C}$, small values lead to lower $Q{}_{i}$, and larger
values lead to higher $Q{}_{i}$. The effect of changing $g{}_{L}$
on $Q{}_{i}$ is at least a factor of three smaller than for $g{}_{C}$.
Thus the gaps where electric fields are present (the capacitor and
not the inductor), partially control $Q{}_{i}$, consistent with a
surface loss mechanism coupled to the electric field. Similarly $Q{}_{i}$
increases for larger $\mathit{w}$, again consistent with surface
loss since wider traces lead to decreasing electric field strength
at surfaces. Next, we find that $Q{}_{i}$ drops by roughly 25\% if
$g{}_{R}$$\geq$50 $\mu$m, suggesting that the ground plane
prevents electric fields from reaching lossy materials such as the
copper box or PCB dielectric. Lastly, the trend indicating that larger
values of $Z{}_{0}$ are beneficial to $Q{}_{i}$ appears to contradict
the usual hypothesis that dissipative mechanisms have a constant $\tan\delta$.
The results for $g{}_{C}$, $g{}_{L}$ and $\mathit{w}$ are all consistent
with a loss dominated by surface electric field participation.

We chose two new sets of parameters from these results with the goal
of improving the $Q{}_{i}$. Resonators with these parameters are
called Design B and Design C resonators. Design B values were chosen
to be relatively modest changes from Design A, while Design C values
were chosen to maximize $Q{}_{i}$. Design B chosen values were: $g{}_{C}$
= 20 $\mu$m, $g{}_{L}$ = 5 $\mu$m, $g{}_{R}$ = 10
$\mu$m, $w$ = 10 $\mu$m and $Z{}_{0}$ = 200 \textgreek{W}.
Resonator size increases rapidly with $g{}_{L}$ since the larger
$Z{}_{0}$ requires twice the inductance. Therefore, to limit the
overall size to roughly 700 $\mu$m x 500 $\mu$m, we
reduced $g{}_{L}$ to 5 $\mu$m, despite the fact that this
may lower $Q{}_{i}$ by 10\%. Design C chosen values were: $g{}_{C}$
= 80 $\mu$m, $g{}_{L}$ = 10 $\mu$m, $g{}_{R}$ =
10 $\mu$m and $Z{}_{0}$ = 300 \textgreek{W}. Note that $g{}_{C}$
was chosen beyond the range of tested mutant Design A resonators.
Also in Design C, the trace width $w$ was different for the capacitor
(40 $\mu$m) and inductor (10 $\mu$m) halves in order
to benefit from the larger capacitor width while keeping the resonator
from being larger than 1000 $\mu$m x 1000 $\mu$m.

\begin{figure}
\begin{centering}
\includegraphics{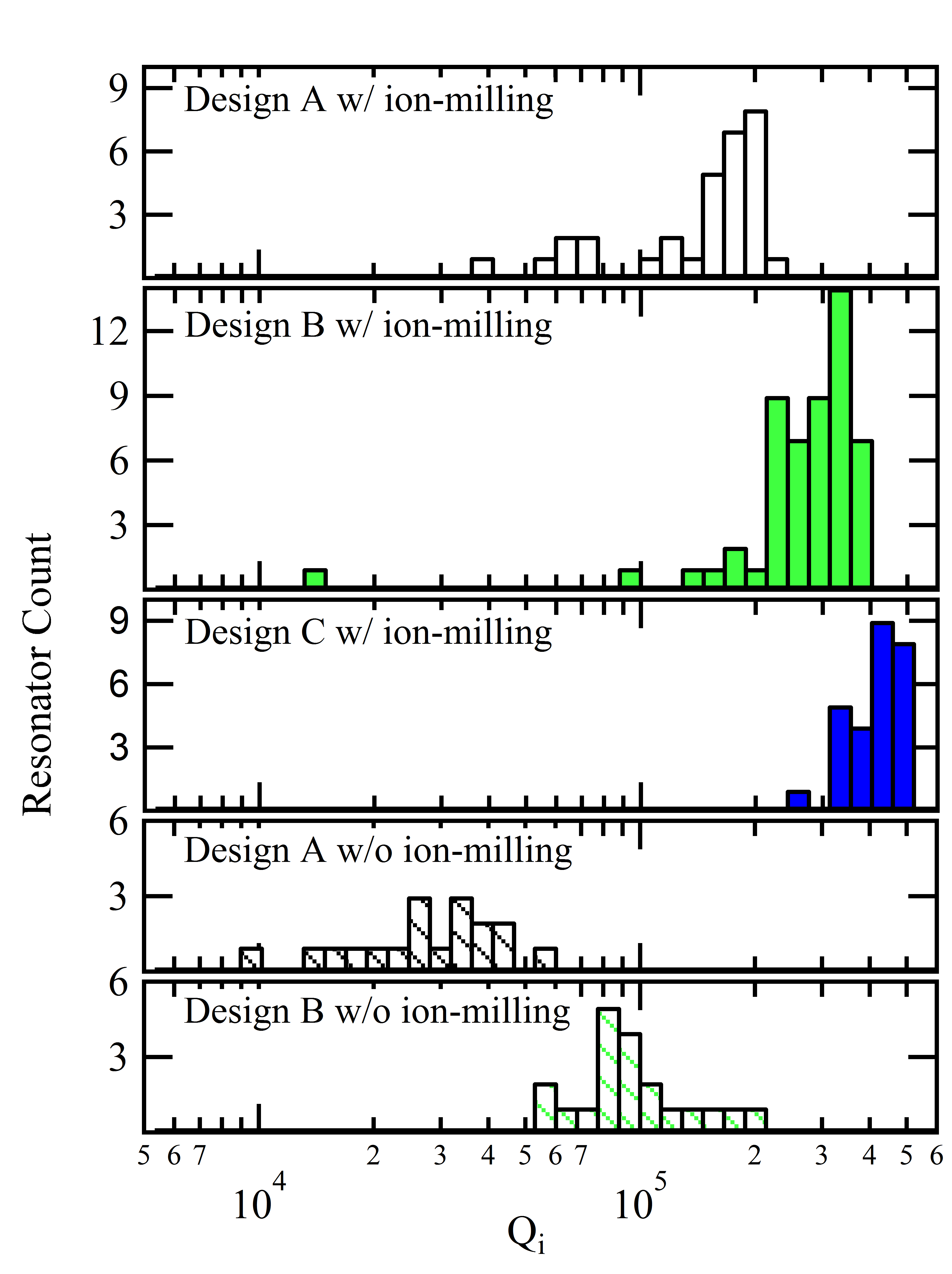}
\par\end{centering}

\caption{\label{fig:Q results}Histograms of single-photon internal quality factors for designs A, B, and C resonators with ion-milling and designs A and B without ion-milling. For both Design A and Design B resonators,
when ion-milling is not performed, $Q{}_{i}$ is roughly a factor
of four lower.}
\end{figure}

The results of all 49 Design A, 73 Design B, and 28 Design C resonators
are shown in Figure \ref{fig:Q results}. Design B and C each show
significantly higher $Q{}_{i}$ than Design A, with Design C on average
better than Design B. While there exists a spread in $Q{}_{i}$ for
each design, we observed an overall increase in the range of measured
$Q{}_{i}$. The maximum/median $Q{}_{i}$ rose from 210,000/160,000
for Design A to 370,000/280,000 for Design B and 500,000/380,000 for
Design C.

When ion-milling was not performed, the maximum/median $Q{}_{i}$
was reduced to 50,000/30,000 for Design A and 190,000/80,000 for Design
B (Design C was not measured without ion-milling). For both Design
A and B, the median quality factor was reduced by roughly a factor
of four when ion-milling was left out during fabrication. Since this
type of cleaning affects only the substrate-air interface and substrate-metal
interface, we infer that these two surfaces participate strongly.
The dominating participation of these surfaces has also been predicted
by simulation \cite{Wenner2011}. This $Q{}_{i}$ dependence on ion-milling
also suggests that while the geometry controls the resonator sensitivity
to the surface loss mechanism, the surface preparation determines
the strength of the loss.

When re-measured in a dilution refrigerator with a lower base temperature
(15mK), we found that resonator $Q{}_{i}$ drops by roughly a factor of 2, which is consistent
with TLS loss \cite{Gao2008}. We have measured similar resonators coupled to a qubit and found that their temperatures would not reach below 50 mK, as also reported by other groups \cite{Corcoles2011}.  However, directly
measuring the linear resonator temperature without a qubit to add nonlinearity is outside the scope of this study. Reassuringly, the increase of $Q{}_{i}$ from Design A to B to C resonators remains even at lower temperatures;
indicating that the geometric variation affects only the sensitivity to loss, not
the absolute strength.

In conclusion, we have shown that the $Q{}_{i}$ of compact resonators
depends strongly on geometrical factors controlling where electric
fields are stored. In addition, substrate surface preparation prior
to metal deposition is crucial. Using our results indicating that
surface loss is dominant, we have been able to increase, at our point of reference temperature of 200 mK, the maximum
internal quality factor of our resonators from 210,000 to 500,000.

The authors thank Danielle Braje at MIT-LL for an independent measurement
of our resonators.  This research was supported by IARPA under grant W911NF-09-1-0369 and ARO under grant W911NF-09-1-0514.


\end{document}